\documentclass[aip, pop, preprint, onecolumn]{revtex4-1}

\newcommand{\be}{\begin{displaymath}}
\newcommand{\bn}{\begin{equation}}

\newcommand{\en}{\end{equation}}
\newcommand{\ee}{\end{displaymath}}

\begin{document}

\title{The universal instability in general geometry}
\affiliation{Max-Planck-Institut f\"ur Plasmaphysik, Wendelsteinstr. 1, 17491 Greifswald, Germany}
\author{P. Helander}
\author{G. G. Plunk}

\begin{abstract}
The ``universal'' instability has recently been revived by Landreman, Antonsen and Dorland \cite{Landreman}, who showed that it indeed exists in plasma geometries with straight (but sheared) magnetic field lines.  Here it is demonstrated analytically that this instability can be present in more general sheared and toroidal geometries. In a torus, the universal instability is shown to be closely related to the trapped-electron mode, although the trapped-electron drive is usually dominant.  However, this drive can be weakened or eliminated, as in the case in stellarators with the maximum-$J$ property, leaving the parallel Landau resonance to drive a residual mode, which is identified as the universal instability.
\end{abstract}

\maketitle

In an ironic turn of events, the ``universal'' instability has recently been revived \cite{Landreman}. This instability of ordinary drift waves was first predicted in the 1960's \cite{Galeev,Krall} but brought into disrepute in the late 1970's by a series of papers demonstrating that it is actually absent in the simplest limit of small ion gyroradius, $k_\perp \rho_i \ll 1$, if there is no temperature gradient or magnetic-field curvature \cite{Antonsen,Mahajan,Catto}. It was even thought (erroneously) that stability prevails at finite $k_\perp \rho_i$, so that a plasma without temperature gradients would always be stable in a simple sheared magnetic field \cite{Chen}. For more information about the long and tortuous history of drift-wave (in)stability, we refer the reader to the reviews by Horton \cite{Horton} and Connor \cite{Connor}. In the latest twist of the story, the recent work of Landreman, Antonsen and Dorland has now shown that the universal instability exists after all, if $k_\perp \rho_i$ is finite. Their (numerical) proof was restricted to a straight, sheared magnetic field, and thus raises the question of whether the instability also exists in more general magnetic geometry. 

The universal instability is electrostatic in nature and requires both the ions and the electrons to be treated kinetically. The dispersion relation is obtained from the gyrokinetic formulation of the quasineutrality condition, which in a hydrogen plasma can be written as
	\bn
	\delta n_i = \delta n_e,
	\label{qn}
	\en
where the perturbed density of species $a$ is
	\bn
	\delta n_a = - \frac{n_a e_a \phi}{T_a} + \int g_a J_0 \mathrm{d}^3 v.
	\en
The perturbed distribution functions $g_a$ are here defined such that the full distribution functions are $f_a = f_{a0}(1 - e_a \phi/T_a) + g_a$, the electrostatic potential is denoted by $\phi$, the argument of the Bessel function is $k_\perp v_\perp / \Omega_a$ (but the electron gyroradius will be neglected), the gyrofrequency is denoted by $\Omega_a = e_a B/m_a$, and the rest of the notation is standard. We consider the instability in the collisionless, electrostatic approximation, where the distribution functions satisfy the gyrokinetic equation
	\bn
	i v_{\|}\nabla_{\|}g_a+(\omega -\omega_{da})g_a 
	= \frac{e_a\phi}{T_a} J_0 \left( \omega - \omega_{\ast a}^T 
	\right)f_{a0}. 
	\label{gk}
	\en
Here, we use the kinetic energy $\varepsilon = m_a v^2/2$ and $\lambda = v_\perp^2 / (v^2 B)$ as independent velocity-space variables, and the parallel derivative $\nabla_{\|}$ is taken holding these constant.  The mode frequency is denoted by $\omega$, the drift frequency by $\omega_{da} = {\bf k} \cdot {\bf v}_{da}$, the diamagnetic frequency by $\omega_{\ast a} = (T_a/n_a e_a B^2) ({\bf k}_\perp \times {\bf B} ) \cdot \nabla n_a $, and we have written
	$$ \omega_{\ast a}^T = \omega_{\ast a} \left[ 1 + \eta_a \left(x^2 - \frac{3}{2} \right) \right], $$
with $x^2 = m_a v^2 / 2 T_a$ and $\eta_a = d \ln T_a / d \ln n_a$. The universal instability is most readily obtained in the limit $v_{Ti}/L_\| \ll \omega$, where $v_{Ti}$ denotes the ion thermal speed and $L_\| = 1/\nabla_\| \ln \phi$ is the length scale of the instability along the magnetic field.  We note that parallel ion motion can be retained perturbatively, as in previous works, but does not significantly alter the subsequent analysis. It is thus neglected here, but details of how it affects the argument are displayed in the Appendix.  For simplicity, we will also neglect $\omega_{di}$; this assumption will be discussed more later.  The solution to the ion gyrokinetic equation is then
	\bn g_i = \frac{e\phi}{T_i} J_0\left( 1 - \frac{\omega_{\ast i}^T}{\omega} \right) f_{i0}, 
	\label{Gi}
	\en
and the quasineutrality condition (\ref{qn}) becomes
	\bn \left[ 1 + \frac{T_e}{T_i} (1-\Gamma_0) - 
	\frac{\omega_{\ast e}}{\omega} \left( \Gamma_0 - \eta_i b \left(\Gamma_0 - \Gamma_1 \right) \right)
	\right] \phi
	+ \frac{T_e}{ne} \int g_e d^3v = 0, 
	\label{integral eq}
	\en
where $\Gamma_n = I_n(b) e^{-b}$ and $b = k_\perp^2 \rho_i^2 = k_\perp^2 T_i / (m_i \Omega_i^2)$. 

What remains is the system composed of equation (\ref{gk}) for the electrons and the constraint (\ref{integral eq}), which is in general difficult to solve analytically.  However, the case of a constant, unsheared magnetic field is relatively simple to analyze, and we examine this case now as it will bring clarity to the general result later.  The electron response in this limit becomes
	$$ g_e = -  \frac{e\phi}{T_e} \frac{\omega - \omega_{\ast e}^T}{\omega - k_\| v_\|} f_{e0}, 
	$$
so that
	\bn \frac{T_e}{ne} \int g_e d^3v = i \sqrt{\pi} \frac{\omega}{k_\| v_{Te}}
	\left[1 - \frac{\omega_{\ast e}}{\omega} 
	\left( 1 - \frac{\eta_e}{2} \right) \right] \phi, 
	\label{Landau resonance contribution}
	\en
if $\omega = \omega_r + i \gamma$ with $\gamma \sim \omega_r \ll k_\| v_{Te}$. It is now a simple matter to find unstable regions in parameter space from Eq.~(\ref{integral eq}). A particularly clear limit is that of small perpendicular wavelength, $k_\perp \rho_i \gg 1$, where 
	$$ \Gamma_0 \simeq \frac{1}{\sqrt{2 \pi b}} \left( 1 + \frac{1}{8b} \right), $$
	$$ \Gamma_1 \simeq \frac{1}{\sqrt{2 \pi b}} \left( 1 - \frac{3}{8b} \right), $$
and thus
	$$ \frac{\omega}{\omega_{\ast e}} \simeq \frac{1}{\sqrt{2 \pi b}} \cdot
	\frac{1 - \eta_i/2}{1 + \frac{T_e}{T_i} - i \sqrt{\pi} \frac{\omega_{\ast e}}{k_\| v_{Te}}
	\left( 1 - \frac{\eta_e}{2} \right)}, $$
implying that the plasma is unstable if $\eta_i$ and $\eta_e$ are both less than $2$ or both greater than $2$. 

We now turn to our main topic of interest, the question of stability in general sheared magnetic geometry.  For topologically toroidal magnetic fields, the gyrokinetic equation as written in Eq.~(\ref{gk}), can be interpreted in ballooning space.  It is worth noting that while previous studies of the universal instability have been done in Fourier space, the formulation in ballooning space is equivalent, as can be demonstrated, with some care, by Fourier transformation.  To pass to ballooning space, we simply write $\nabla_{\|} = \partial/\partial l$, where $l$ denotes the arc length along the magnetic field line, now taken to extend infinitely in both directions.  As is well known, solutions in the torus are obtained by transformation of the ballooning space solutions, but we note that one can also interpret solutions as applying directly to infinite sheared geometries, of which the simplest example is the sheared slab.  The following analysis is to be interpreted in both ways, {\em i.e.} as applying to the old sheared-slab universal instability, and also as applying to a generalized toroidal version.

Because the magnetic field strength and the shear vary along the magnetic field, the mode structure and the dispersion relation can no longer be obtained analytically, but we can nevertheless extract most of the information we need by the quadratic form obtained by multiplying Eq.~(\ref{integral eq}) by the complex conjugate of the electrostatic potential, $\phi^\ast$, and integrating along the entire field line, $-\infty < l < \infty$, 
	\bn \int_{-\infty}^\infty 
	\left[ 1 + \frac{T_e}{T_i} (1-\Gamma_0) - 
	\frac{\omega_{\ast e}}{\omega} \left( \Gamma_0 - \eta_i b \left(\Gamma_0 - \Gamma_1 \right) \right)
	\right] | \phi |^2 \frac{dl}{B} + \frac{T_e}{ne} \int_{-\infty}^\infty \phi^\ast \frac{dl}{B}
	\int g_e d^3v = 0. 
	\label{torus}
	\en
The imaginary part of this equation is
	\bn \frac{\gamma \omega_{\ast e}}{\omega_r^2 + \gamma^2} 
	\int_{-\infty}^\infty 
	\left[ \Gamma_0 - \eta_i b \left(\Gamma_0 - \Gamma_1 \right) \right]| \phi |^2 \frac{dl}{B}
	= \frac{T_e}{ne^2} Q_e(\omega), 
	\label{gamma}
	\en
where
	$$ Q_e(\omega) = - e \; {\rm Im} \; \int_{-\infty}^\infty \phi^\ast \frac{dl}{B}
	\int g_e d^3v, $$
and we conclude that instability is impossible unless $Q_e(\omega)/\omega_{\ast e}$ is positive. Physically, $Q_e$ is proportional to the work done by the electrons on the instability \cite{Horton,Proll,Helander}. It was calculated in Ref.~\cite{Proll} and found to be a sum of contributions from passing and trapped particles,
	$$ Q_e(\omega) = Q_{ep}(\omega) + Q_{et}(\omega), $$
where, if $0 < \gamma \ll |\omega_r|$, the former is given by 
\bn
Q_{ep}(\omega)=-\frac{n e^2}{T_e v_{Te}\sqrt{\pi}}\int_0^{\infty}(\omega -\omega^T_{*e}) e^{-x^2}x\mathrm{d}x
\int_{0}^{1/B_{max}}\mathrm{d}\lambda \sum_{j=\cos, \sin} \left| \psi_j(x,\lambda,\omega)  \right|^2,
\label{Qap}
\en
with 
	$$ {\psi_{\cos}(x,\lambda, t) \choose \psi_{\sin}(x,\lambda,t)}
	=\int_{-\infty}^{\infty} {\cos \choose \sin} M(t,0,l)
	\frac{\phi(l) \mathrm{d}l}{\sqrt{1-\lambda B}} , $$
and 
	$$ M(\omega,a,b)=\int_{a}^{b}\left(\omega-\omega_{da}\right)
	\frac{\mathrm{d}l'}{\left|v_{\|}\right|}, $$
The corresponding contribution from the trapped particles is given by 
$$
Q_{et}(\omega)=-\frac{2 \sqrt{\pi} n e^2}{T_e v_{Te}}\sum_{m=-\infty}^{\infty}
\int_0^{\infty} (\omega -\omega^T_{*e}) e^{-x^2}x\mathrm{d}x $$
\bn \times
\int_{1/B_{min}}^{1/B_{max}}\mathrm{d}\lambda \sum_{\mathrm{wells}}
\delta \left( M(\omega,l_1, l_2)-m\pi \right)\left|\psi_t(x,\lambda,\omega) \right| ^2,
\label{Qat}
\en
where the sum is taken over all the trapping wells along the field line, $l_{1,2}$ denote the locations of consecutive bounce points, and 
 $$
\psi_t(x,\lambda,\omega)
= \int_{l_1}^{l_2}\frac{\cos M(\omega,l_1, l)}{\sqrt{1-\lambda B}} \; \phi(l) \mathrm{d}l.
$$

We now take the limit $\omega \ll k_\| v_{Te}$, where $M \ll 1$, and note that the passing-particle contribution (\ref{Qap}),
	\bn \frac{Q_{ep}(\omega)}{\omega_{\ast e}} = \frac{ne^2}{2 \sqrt{\pi} v_{Te} T_e}
	\left( 1 - \frac{\eta_e}{2} - \frac{\omega}{\omega_{\ast e}} \right) 
	\int_0^{1/B_{\rm max}} \left| \psi_{\rm cos} \right|^2 d \lambda, 
	\label{Qep}
	\en
exactly corresponds to that producing the universal instability in a straight magnetic field, Eq.~(\ref{Landau resonance contribution}), but is relatively small compared with the contribution from trapped electrons. The latter is dominated by the $m=0$ term in the sum (\ref{Qat}), and because of the delta function, this term picks out the resonance $\omega = \overline \omega_{de}$ (an overbar denotes the bounce average) and is larger than Eq.~(\ref{Qep}) by a factor $k_\| v_{Te} / \omega \gg 1$.

We are thus led to the conclusion that in arbitrary toroidal geometry the same driving mechanism (parallel Landau resonance) is present to drive an instability as in the case of a straight magnetic field, but it is generally overwhelmed by the trapped-particle drive. To prove that an instability similar to the universal one nevertheless exists in certain geometries, we consider the limit of steep density gradient, $\omega_{\ast e} \gg \omega_{de}$. In this limit, we can neglect $\omega_{de}$ in a first approximation, so that the trapped-electron response simply becomes
	$$ g_e = -  \left(1 - \frac{\omega_{\ast e}}{\omega} \right) 
	\frac{e \overline \phi}{T_e} f_{e0}. $$
Equation (\ref{torus}) then yields the following variational form for the real frequency \cite{Helander},
		\bn \frac{\omega}{\omega_{*e}} = \frac{N[\phi]}{D[\phi]}, 
	\label{variational form}
	\en
where the functionals $N$ and $D$ are defined by 
	$$ N[\phi] = \int_{-\infty}^\infty \Gamma_0 |\phi |^2 \frac{dl}{B} 
	- \frac{1}{2} \int_{1/B_{\rm max}}^{1/B_{\rm min}} \sum_j \tau_j | \overline{\phi}_j |^2 d\lambda, $$
	$$ D[\phi] = \int_{-\infty}^\infty \left[ 1 + \frac{T_e}{T_i} (1 - \Gamma_0) \right] |\phi |^2 \frac{dl}{B} 
	- \frac{1}{2} \int_{1/B_{\rm max}}^{1/B_{\rm min}} \sum_j \tau_j | \overline{\phi}_j |^2 d\lambda. $$
and we have taken $\eta_i = 0$ for simplicity. The sums appearing in these expressions are taken over all trapping wells along the field line, and 
	$$ \tau_j(\lambda) = \int_{l_1}^{l_2} \frac{dl}{\sqrt{1 - \lambda B(l)}} $$
denotes the normalized bounce time in such a well. The key idea here is that Eq.~(\ref{variational form}) furnishes the zeroth order solution ($\omega_r$, $\phi(l)$).  The growth rate is then computed directly from Eq.~(\ref{gamma}), which exactly accounts for the resonances.  We do not do this computation explicitly, but argue for the existence of a particular solution as follows.  

The denominator $D[\phi]$ in Eq.~(\ref{variational form}) is always positive, because of the Schwartz inequality, 
$$ \frac{1}{2} \int_{1/B_{\rm max}}^{1/B_{\rm min}} \sum_j \tau_j | \overline{\phi}_j |^2 d\lambda
	\le \sqrt{1-\frac{B_{\rm min}}{B_{\rm max}}} \int |\phi |^2 \frac{dl}{B}, $$
and so is the numerator $N[\phi]$ when the perpendicular wavelength $b = k_\perp^2 \rho_i^2$ vanishes. However, when $b \rightarrow \infty$, $N[\phi]$ becomes negative and must therefore vanish at some intermediate value of $b$. There is, therefore, a range of perpendicular wavelengths for which the real frequency satisfies $0 < \omega / \omega_{\ast e} < 1$, making the passing-electron contribution to the growth rate (\ref{Qep}) positive if $\eta_e < 2$, regardless of the (unknown) mode structure $\phi(l)$. As already remarked, this contribution is however usually overwhelmed by the trapped-particle drive (\ref{Qat}), but the latter can be made arbitrarily small by tailoring the magnetic field. In a tokamak, this can be achieved by increasing the aspect ratio so as to reduce the fraction of trapped particles. In a stellarator, it can be accomplished by minimizing the amount of unfavorable bounce-averaged curvature. In the limit of a so-called maximum-$J$ stellarator \cite{Proll,Helander}, all trapped particles experience average favorable curvature, so that $\omega_{\ast e} \overline \omega_{de} < 0$ for all orbits. Because $\omega$ has the same sign as $\omega_{\ast e}$, there are then no resonances in the delta function of Eq.~(\ref{Qat}) and the trapped-electron drive vanishes. In both cases, we are left with an instability driven by circulating electrons through a parallel Landau resonance producing a response (\ref{Qep}) very much like that in a straight magnetic field. Note that this instability does not require a temperature gradient but can be driven by a density gradient alone. Finally, for the purposes of comparing this result with previous work, we note that inclusion of finite parallel ion motion in the ion response merely introduces an extra quadratic term to the left-hand-side of Eq.~(\ref{torus}), as shown in the Appendix. This term is of order $(k_\| v_{Ti}/\omega)^2$ compared with the others, and is additionally small at large $k_\perp \rho_i$. 

Let us summarize and discuss our findings.  First, we have given new theoretical support for the existence of the universal instability in a sheared slab geometry, the context in which the mode has been traditionally studied.  This may not at once be apparent, but it is because the sheared slab is a limit of the gyrokinetic system in ballooning space, and this limit can be approached as described above, by continuously deforming a general magnetic field.  Our argument requires $k_\perp \rho_i = O(1)$, and so does not contradict stability proofs \cite{Antonsen, Mahajan, Catto} where this parameter was taken to be small.  We have also established the existence of the instability in topologically toroidal magnetic geometry.  In particular, maximum-$J$ stellarators prove an intriguing example:  It turns out that the stability of trapped-electron modes in these configurations leaves space for the  universal instability to appear.  It seems suitably ironic that the instability called universal may ultimately be found lurking in the most exotic of places.

\section*{Appendix: Parallel ion motion}

When the first term in Eq.~(\ref{gk}) is small for the ions, it can be accounted for perturbatively, by denoting the lowest-order solution (\ref{Gi}) by $G_i$ and writing
	$$ g_i = G_i - \frac{i v_\|}{\omega} \nabla_\| 
	\left( G_i - \frac{i v_\|}{\omega} \nabla_\| g_i \right). $$
Approximating the last term on the right by $G_i$, we find the ion density perturbation to be
	$$ \frac{\delta n_i}{n} = - \frac{e \phi}{T_i} + \frac{\delta n_0 + \delta n_1}{n} , $$
with
	$$ \frac{\delta n_0}{n} = \frac{1}{n} \int G_i J_0 d^3v = \frac{e \phi}{T_i} 
	\left[ -1 + \left( 1 - \frac{\omega_{\ast i}}{\omega} \right) \Gamma_0
	- \frac{\eta_i \omega_{\ast i}}{\omega} b (\Gamma_0 - \Gamma_1 ) \right], $$
	$$ \frac{\delta n_1}{n} = - \frac{B}{n \omega^2} \int_0^\infty f_{i0} 2 \pi v^4 dv
	\int_0^{1/B} J_0 \nabla_\| \left[ \sqrt{1-\lambda B} \; \nabla_\| (J_0 \phi) \right] d\lambda, $$
and in Eq.~(\ref{torus}) we thus obtain the following additional term on the left-hand side
	$$ \frac{T_e}{n \omega^2 T_i}
	\int_{-\infty}^\infty \frac{dl}{B} \int f_{i0} v_\|^2 v^2 \left| \nabla_\| (J_0 \phi) \right|^2
	d^3v, $$
which is a small correction, of order $(k_\| v_{Ti} / \omega)^2$, to the other terms.

\end{document}